\documentclass[aps,prl,reprint,superscriptaddress,amsmath,amssymb,longbibliography]{revtex4-1}

\usepackage{graphicx}
\usepackage{bm}
\usepackage[colorlinks=false]{hyperref}
\usepackage{color,soul}
\usepackage{todonotes}

\begin{document}


\title{Direct Measurements of Collisional Dynamics in Cold Atom Triads}


\author{L. A. Reynolds}
\affiliation{The Dodd-Walls Centre for Photonic and Quantum Technologies, New Zealand}
\affiliation{Department of Physics, University of Otago, Dunedin, New Zealand}
\author{E. Schwartz}
\affiliation{The Dodd-Walls Centre for Photonic and Quantum Technologies, New Zealand}
\affiliation{Department of Physics, University of Otago, Dunedin, New Zealand}
\author{U. Ebling}
\affiliation{The Dodd-Walls Centre for Photonic and Quantum Technologies, New Zealand}
\affiliation{Centre for Theoretical Chemistry and Physics, New Zealand Institute for Advanced Study, Massey University,  Auckland, New Zealand}
\author{M. Weyland}
\affiliation{The Dodd-Walls Centre for Photonic and Quantum Technologies, New Zealand}
\affiliation{Department of Physics, University of Otago, Dunedin, New Zealand}
\author{J. Brand}
\affiliation{The Dodd-Walls Centre for Photonic and Quantum Technologies, New Zealand}
\affiliation{Centre for Theoretical Chemistry and Physics, New Zealand Institute for Advanced Study, Massey University,  Auckland, New Zealand}
\author{M. F. Andersen}
\email[Email: ]{mikkel.andersen@otago.ac.nz}
\affiliation{The Dodd-Walls Centre for Photonic and Quantum Technologies, New Zealand}
\affiliation{Department of Physics, University of Otago, Dunedin, New Zealand}


\date{\today}

\begin{abstract}
The introduction of optical tweezers for trapping atoms has opened remarkable opportunities for manipulating few--body systems. Here, we present the first bottom--up assembly of atom triads. We directly observe atom loss through inelastic collisions at the single event level, overcoming the substantial challenge in many--atom experiments of distinguishing one-, two-, and three--particle processes. We measure a strong suppression of three--body loss, which 
is not fully explained by the presently availably theory for three-body processes. 
The suppression of losses could indicate the presence of local anti--correlations 
due to the interplay of attractive short range interactions and low dimensional confinement.
Our methodology opens a promising pathway in experimental few--body dynamics.
\end{abstract}

\maketitle


An enduring ambition in atomic physics is to build an understanding of interacting macroscopic systems entirely from knowledge of the underlying microscopic dynamics. In recent years, experimental advancements in isolation and control of single atoms \cite{Gruenzweig2010, Isenhower2010, Kuhr2001, Thompson2013, Rosnagel2016, Barredo2016, Endres2016, Goban2018} paved the way for connecting the few--body and many--body regimes \cite{Serwane336}. In particular, optical dipole traps (optical tweezers) proved instrumental in demonstrations of fundamental atomic phenomena like molecular formation and inelastic collisions \cite{Sompet2013, Sompet2019, Xu2015, Liu2018, Lester2018}. Conversely, large atomic samples such as Bose--Einstein condensates (BECs) provide a tool for studying atomic dynamics from the many-body perspective \cite{Cornish2006, Kraemer2006, Levinsen2015, Klauss2017}. Nonetheless, dynamics of large samples are complex with many processes affecting the observed signals simultaneously.

While many phenomena observed in BECs are accurately described in a mean--field framework, the loss processes induced by inelastic particle collisions are strongly influenced  by correlations that are omitted in this description \cite{Burt1997}.
Measured inelastic collision rates therefore provide invaluable information about correlations in a system. Strongly confined repulsive BECs may undergo fermionization with suppressed local correlations leading to reduced atom loss \cite{Gangardt2003a, Gangardt2003b, Tolra2004}. On the contrary, attractive interactions typically lead to a collapse of the condensate \cite{Kagan1998, Roberts2001, Bradley1997}, soliton formation \cite{Cornish2006, Strecker2002, Nguyen2017, Donley2001}, or few-body (Efimov) bound states or resonances \cite{Efimov1970,Kraemer2006,Ferlaino2011} with enhanced \emph{three-body} loss rates. Three--body recombination happens when three atoms approach within their interaction range, two atoms form a molecule, and the third receives a share of the released binding energy. The process is sensitive to three--particle correlations \cite{Gangardt2003a, Gangardt2003b} and has interesting consequences for the many-body dynamics \cite{Jack2002}, while the accurate modelling of the recombination process is a huge challenge \cite{Esry1999a}.
Three--body recombination occurs throughout physics from ultracold plasmas \cite{Lyon2016} to chemistry \cite{Baulch1992} and astrophysics \cite{Forrey2013} and has been extensively studied in ultracold atoms \cite{Weber2003, Tolra2004, Gross2009, Braaten2006, DIncao2015, Roberts2000,Altin2011}. 
Moreover, the rich physics of idealised three atom systems in tightly confining traps is currently the target of intensive theoretical studies \cite{Guijarro2018, Nishida2018, Pricoupenko2018, Pricoupenko2019, Valiente2019, Happ2019}, while experiments are presently lacking.

Here we report the first controlled fabrication and manipulation of atom triads via a bottom--up approach of assembling atomic samples. By isolating three independent $^{85}$Rb atoms in separate optical tweezers and dynamically bringing them together, we obtain the first experimental observation of distinguishable few--atom inelastic collisions at the single event level. We find a strongly suppressed three--body recombination loss rate compared to previous experiments with many--atom ensembles
\cite{Roberts2000,Altin2011,Wild2012}. 
There is currently no reliable theory for quantitatively describing three-body processes in 
an optical tweezer trap. 
While resonant three-body physics or a modification of the three-body process itself 
could be relevant, we argue that the suppressed loss rate may indicate 
the presence of anti-correlations similar to those present in the super-Tonks-Girardeau gas, a metastable phase of attractively interacting bosons in one dimension \cite{Astrakharchik2005,Haller2009}. We also measure an increased two--body loss which we attribute to  photo-assisted processes due to the dipole trap laser field.

\begin{figure}[h]
\includegraphics[clip, trim=1.95cm 3.55cm 3.5cm 11cm, width=0.5\textwidth]{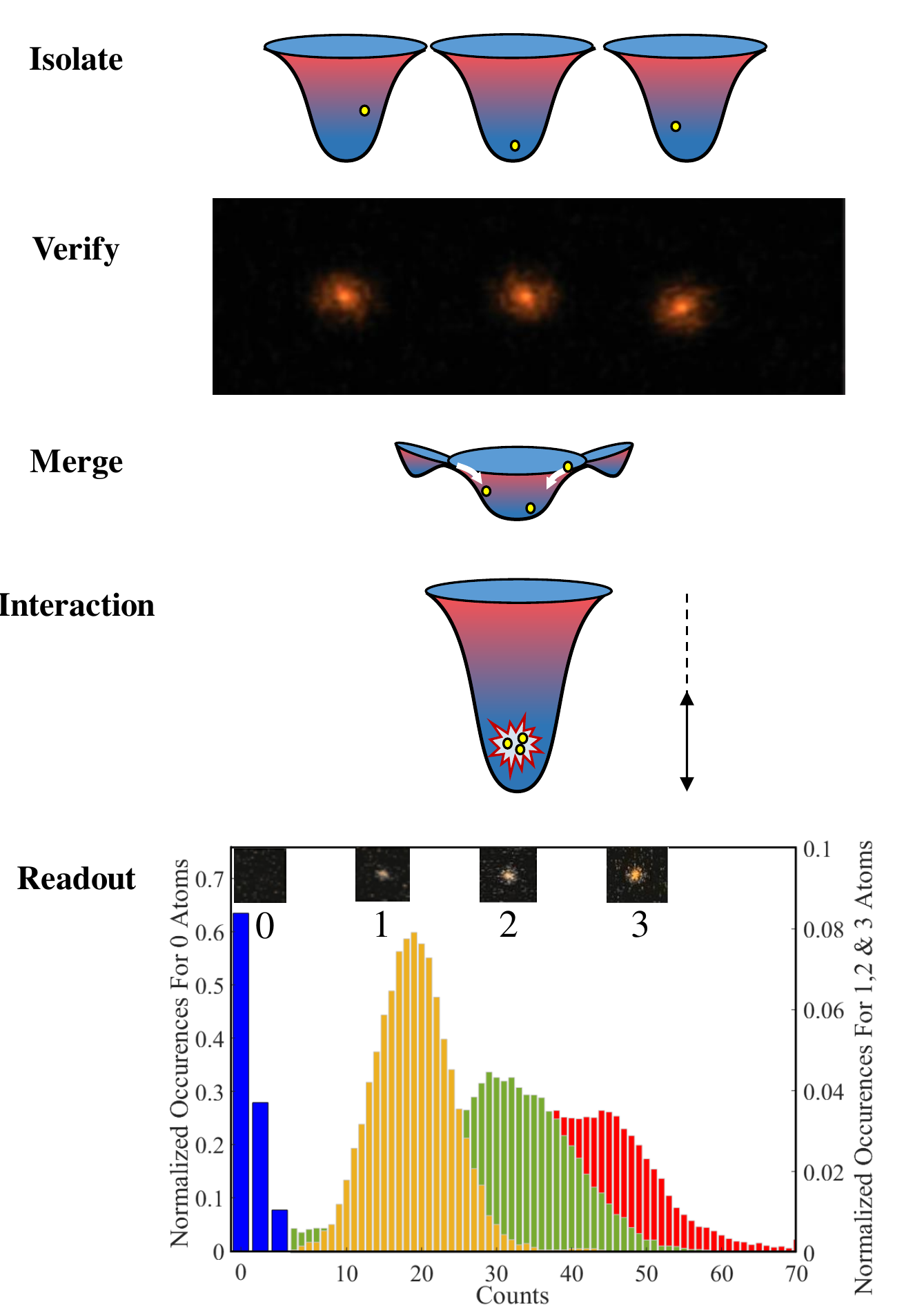}%
\caption{\label{mergefig}(Color online) Experimental procedure for directly observing cold atom collisions. We isolate three \textsuperscript{85}Rb atoms in separate optical tweezers and confirm their presence through fluorescence imaging.  A merge and compression stage allows the atoms to interact. Readout: number of photon occurrences for zero (blue), one (yellow), two (green), and three (red) atoms in the tweezer}
\end{figure}

Figure \ref{mergefig} portrays the experimental process. It starts by isolating three atoms in three optical tweezers separated by $\sim4.5\,\mu$m using a similar method as used for two atoms in \cite{Sompet2019}. The isolation stage utilizes blue--detuned light--assisted collisions yielding a single atom in each tweezer with high probability \cite{Carpentier2013, Fung_2015, Brown2019}. A high-numerical-aperture lens (NA = 0.55) focuses three steerable linearly polarized laser beams ($\lambda=1064$ nm) to a spot size of $\omega_0 =1.1$ $\mu$m to form the tweezers. A fluorescence image confirms the presence of the three isolated atoms \cite{Hilliard2015, Sompet2019}. The trap oscillation frequencies (measured by Raman sideband spectroscopy) are $\{210, 210, 34\}$ kHz for 110 mW beam power \cite{Sompet2017}. For other beam powers, the trap frequencies scale with the square root of the power. 
Before merging, a $\sigma_{-}$ polarized light beam addressing the $D_1$ $|F=2\rangle$ to $|F'=2\rangle$ transition prepares the atoms in the $|F=2, m_{F}=-2\rangle$ ground state with 99.1\% efficiency. During the preparation, a re--pump beam on the $D_2$ $|F=3\rangle$ to $|F'=3\rangle$ transition prevents population buildup in the $|F=3\rangle$ state, and a bias magnetic field of 8.5 G defines the quantization axis.

Using an acousto--optic modulator (AOM), we steer the two outer tweezers closer to the central one until they all merge. The duration of this manipulation is 40 ms, and the final spacing, in merged state, between the centers of the three beams is 0.85 $\mu$m. After ramping off the two outermost tweezers adiabatically in $\sim$30 ms, the single tweezer beam, holding three atoms, ramps adiabatically from 5 mW to 110, 140, 170, or 200 mW. The samples' peak density range is then $0.9-1.5\times10^{14}$ atoms/cm$^{-3}$. The atoms collide for varying controlled time duration (denoted 'wait time') before the remaining population is determined by using a single photon counting module to detect fluorescence \cite{McGovern2012, Hilliard2015}. The 'Readout' section of Fig.\ \ref{mergefig} shows the photon distributions for 0 (blue), 1 (yellow) 2 (green), or 3 (red) atoms in the single tweezer. The distributions are not entirely separated, but they are sufficiently distinct to allow determination of the probability distribution for each atom number. We fit a weighted sum of them to the measured photon distribution for each wait time. Each photon distribution is sampled from at least 600 experimental repetitions.  

The ensemble temperature after merging is 17.8 $\mu$K with a tweezer beam power of 5 mW as determined via the release--and--recapture (RR) technique \cite{Tuchendler2008}. The temperature scales as the square root of the trap beam power, which allows us to infer the temperature at the `collision depth'. We verified that the ramp to the collision depth was adiabatic by seeing no significant difference between the temperature before the ramp and after ramping up and back down.

To model atom loss dynamics we use the theory of open quantum systems in the Born--Markov approximation.
This is adequate if the atoms are lost from the trap in processes that happen expeditiously relative to the time scale of in--trap dynamics \cite{Jack2002}. The Born--Markov master equation for the density operator $\hat\rho$ is:
\begin{align}
\label{eq:MasterEquation}
    \frac{d\hat\rho}{dt}=&-\frac{i}{\hbar}\left[\hat H_T,\hat\rho\right]+\sum_{j=1}^3\kappa_j\int d^3r\left[2\hat\psi^j(\mathbf{r})\hat\rho\hat\psi^{\dagger j}(\mathbf{r})\right.\nonumber\\
    &\left. -\hat\psi^{\dagger j}(\mathbf{r})\hat\psi^j(\mathbf{r})\hat\rho-\hat\rho\hat\psi^{\dagger j}(\mathbf{r})\hat\psi^j(\mathbf{r})\right],
\end{align}
where the coefficients $\kappa_j$ describe the strength of the $j$-body loss processes
and $\hat H_T$ describes the conservative dynamics of the atoms in the trap. When only three or fewer particles are present in the trap,
we may derive 
a set of rate equations for the probabilities $r_i(t)$ for observing $i$ atoms in the tweezer at a given time \cite{SuppMat}:
\begin{align}
\label{decayeqns}
\begin{split}
   \dot r_3(t)&=-\Gamma_3 r_3- 3\widetilde\Gamma_2 r_3 - 3 \Gamma_1 r_3 ,\\
   \dot r_2(t)&=-\Gamma_2 r_2- 2\Gamma_1 r_2 + 3 \Gamma_1 r_3 ,\\
   \dot r_1(t)&=-\Gamma_1 r_1+ 3\widetilde\Gamma_2  r_3 + 2 \Gamma_1 r_2 ,\\
   \dot r_0(t)&=\Gamma_3 r_3+\Gamma_2 r_2+\Gamma_1 r_1 .
\end{split}
\end{align}
where  $\Gamma_1=2\kappa_1$.  
The rate coefficients  $\Gamma_j$ for two- and three--body losses depend on integrated local $j$-body correlation functions for $N$ particles,
\begin{align} \label{eq:corrfunc}
    \mathcal{C}_N^j\equiv \int d^3r\langle\hat\psi^{\dagger j}(\mathbf r)\hat\psi^j(\mathbf r)\rangle_N 
\end{align}
and specifically, 
\begin{align}
\label{eq:gamma3}
   \Gamma_3=2\kappa_3C^3_3,
   \quad \Gamma_2=2\kappa_2C^2_2, \quad \widetilde\Gamma_2=\frac{2}{3}\kappa_2C^2_3.
\end{align}
Note that when the three--body coefficient $\kappa_3$ is known, a measurement of the rate coefficient $\Gamma_3$ constitutes a measurement of the three--body correlations in the sample. 
$\kappa_3$ can be obtained  from measurements or theoretical calculations of the three--body recombination rate coefficient $K_3$. For the purpose of interpreting our results we use $\kappa_3=0.093\times 10^{-25}\, \text{cm}^6/\text{s}$, which corresponds to the value of $K_3$ in Ref.~\cite{Esry1999a}.

When working with individually assembled triads, the dynamics cease as soon as there is a loss event. This eliminates the need to model a changing density profile as within a large ensemble. We therefore assume the linear rate constants $\Gamma_j$ to be time independent and extract them from fitting experimental data to the solutions of Eqs.~\eqref{decayeqns} \cite{SuppMat}.



\begin{figure}[h]
\includegraphics[clip, trim=2.2cm 8.2cm 2cm 7.32cm,width=0.48\textwidth]{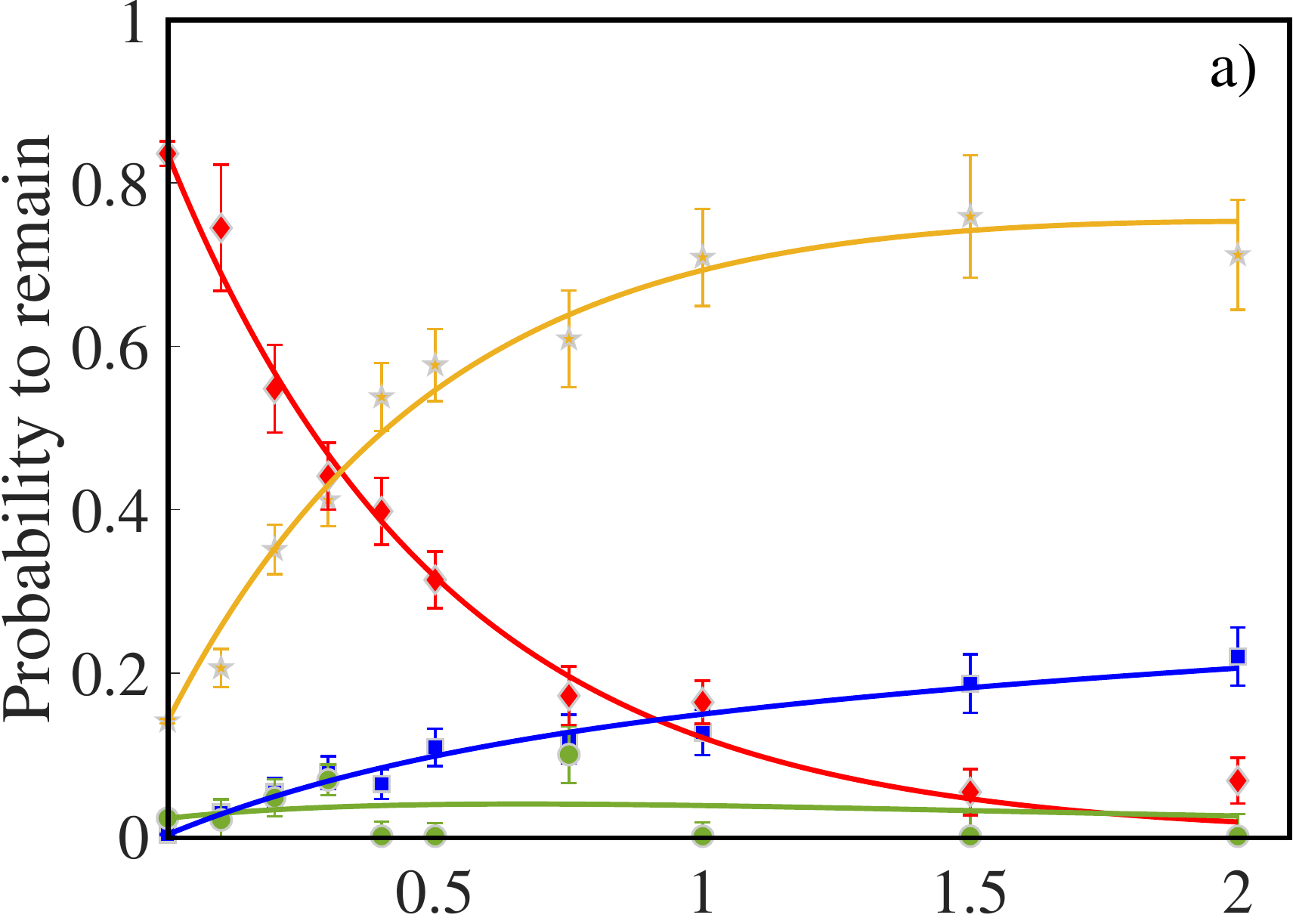}\\
\includegraphics[clip, trim=2.2cm 7cm 2cm 7.2cm,width=0.48\textwidth]{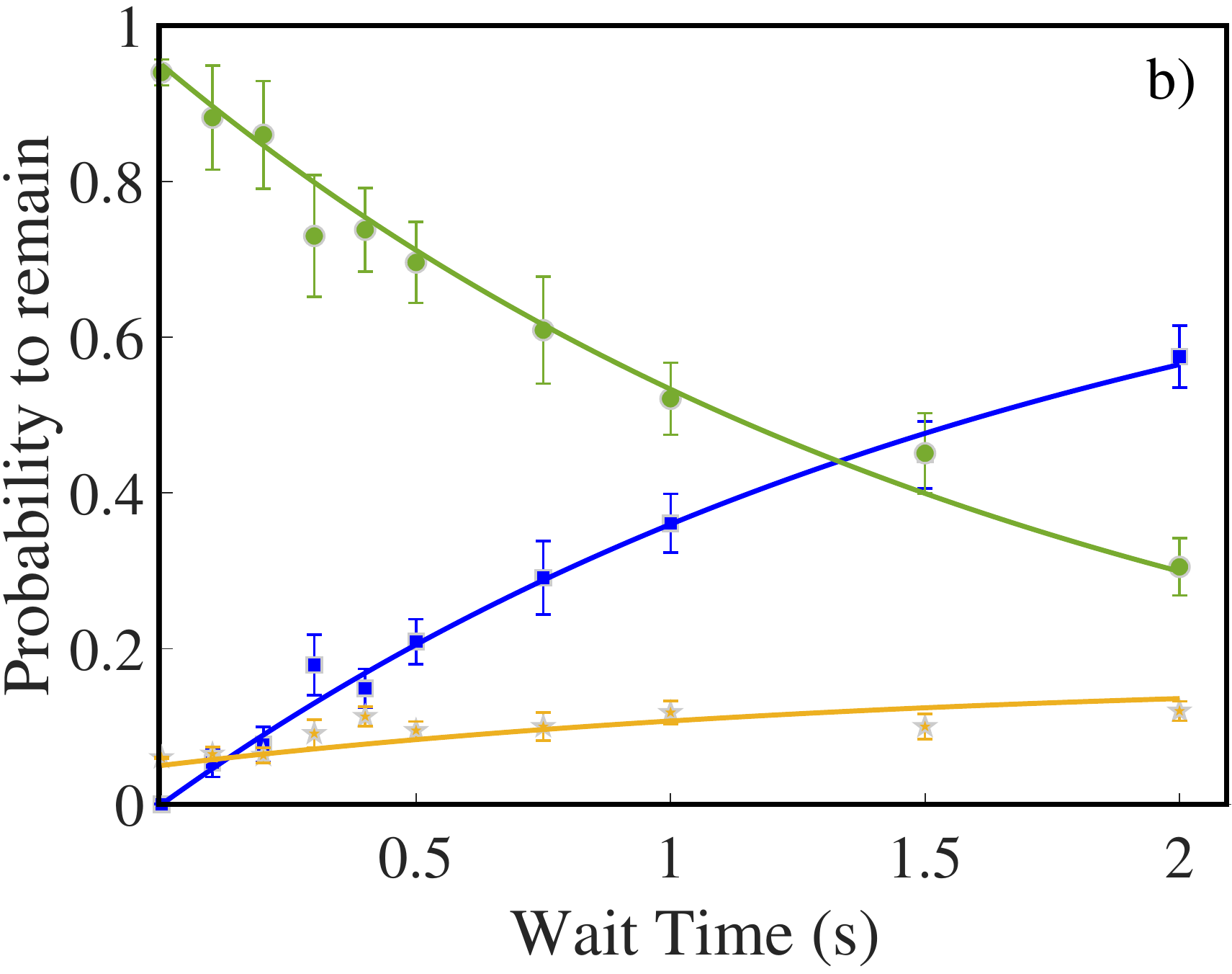}%
\caption{\label{decayplot}Measurements of loss dynamics in triads (a) and dyads (b). Measured probability for the remaining atom number being: Three, Red/Diamond; Two, Green/Circle; One, Yellow/Star; Zero Blue/Squares. Solid lines signify a fit to the data with proper $\Gamma_j$ for triads (Eq. \ref{decayeqns}) and dyads.}
\end{figure}

Figure \ref{decayplot} presents example plots of the population dynamics in a tweezer with a beam power of 170 mW.
When the probability of observing three atoms in Fig.\ \ref{decayplot}a) (red diamonds) decays, the probabilities for observing one (yellow stars) or zero (blue squares) atoms grows. The probability for observing two atoms remains effectively zero for all times, showing that single-atom loss is negligible in the experiment. The data directly reveal whether a loss event is a three--body event that leads to zero atoms remaining or a two--body event that leads to one atom remaining. The solid lines represent a fit with the solutions to Eq.\ \eqref{decayeqns} with $r_j\left( 0 \right)$ and $\Gamma_j$ free parameters.

\begin{figure}[h!]
\centering
\includegraphics[clip, trim=5.9cm 9.95cm 5.5cm 10cm,width=0.5\textwidth]{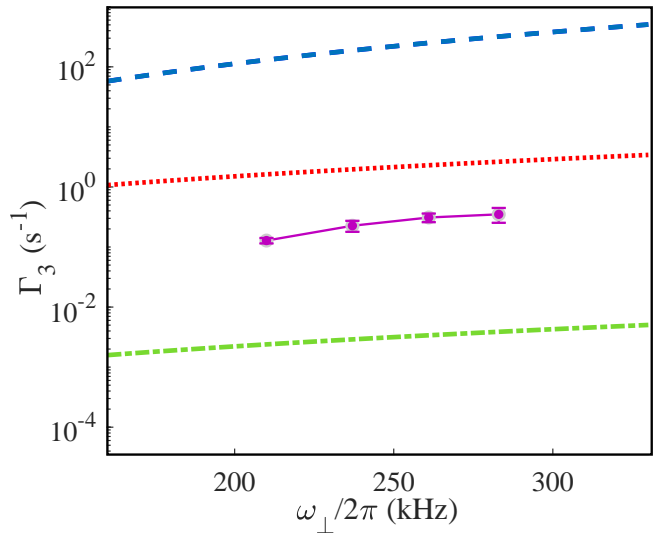}%
\caption{\label{gamma3theory} Comparison of experimentally measured values of the three-body loss rate $\Gamma_3$ (purple/solid) with different theoretical models. Blue/dashed: a thermal gas without interaction--induced correlations. Green/dash--dotted: super-Tonks-Girardeau correlations with an integrated thermal density. Red/dotted: Assumes a 1D gas with no occupation of transverse excited modes. 
}
\end{figure}



Figure~\ref{gamma3theory} shows how the measured $\Gamma_3$ coefficient (purple circles) varies with transverse trap oscillation frequency. %
We compute $C^3_3$ for a thermal gas without interaction--induced correlations as
$^\mathrm{th}C^3_3=\frac{4}{3}\int \mathrm{d}^3r\: n^3(\mathbf{r})$, where the density, $n(\mathbf{r})$, is approximated semi--classically as, $n(\mathbf{r}) = n_0 e^{-V(\mathbf{r})/kT}$,  and $V(\mathbf{r})$ is the optical trapping potential. This scenario is considered in previous works \cite{Roberts2000,Esry1999a}, with the exception of the pre--factor, which is specific to a three--particle system \cite{SuppMat}. The blue dashed line in Fig~\ref{gamma3theory} shows that this prediction lies significantly higher than the measured three--body recombination rate. 
Equation (\ref{eq:gamma3}) indicates that a natural candidate for explaining the observed suppression of three--body loss is interaction--induced anti--correlations. Strong anti--correlations are common in one-dimensional (1D) scenarios for both elastic \cite{Tolra2004} and inelastic \cite{Syassen2008,Daley2009} interactions, and our tweezer has a high aspect ratio with $\omega_z/\omega_\perp\approx 0.16$. $^{85}$Rb's elastic interaction is attractive (negative scattering length, $a=-475a_{0}$ in this system \cite{Claussen2003}), where naively the opposite effect is expected. However, anti--correlations occur in the super--Tonks--Girardeau gas, which is an excited state of a one-dimensional (1D) attractive Bose gas \cite{Astrakharchik2005,Haller2009}, and originate from unrealized two-particle bound states causing an excluded volume similar to the case of hard spheres. 
In the experiment, a finite temperature of $k_{B}T \approx 51.37 \; \hbar\omega_z$  gives a low statistical weight to the
two- and three-particle bound states (solitons),
and the situation could be similar to the super--Tonks--Girardeau gas. 

To check if super--Tonks--Girardeau--like anti--correlations could be responsible for the three--body loss suppression, we estimate the maximally feasible suppression due to elastic two--body scattering, only. Writing $C^3_3\approx g_3\,^\mathrm{th}C^3_3$, the correlation factor  $g_3={\langle\hat\psi(z)^{\dagger 3}\hat\psi(z)^3 \rangle}/{n(z)^3}$, defined for a 1D gas, is $g_3\approx \frac{16 \pi^6}{15 \gamma_\mathrm{LL}^6} $ \cite{Gangardt2003a,Kormos2011}, where $\gamma_\mathrm{LL} = 2a/[n_{1D}(0)l_\perp^2(1-C\tfrac{a}{l_\perp})]$ is the dimensionless Lieb-Liniger constant which is a 1D coupling constant with the transverse oscillator length defined as $l_\perp=\sqrt{\hbar/{m\omega_\perp}}$ ($\mathcal{O}\; 100$nm) and $C=1.0326\ldots$\cite{Astrakharchik2005}. Taking the maximal transversely integrated particle density for $n_{1D}$  gives the green/dash--dotted line in Fig.~\ref{gamma3theory} and $g_3\approx 10^{-5}$. This lower bound on $\Gamma_3$ shows that super--Tonks--Girardeau--like anti--correlations could be responsible for the suppression.

At our experimental temperature, the sample is not in the transverse ground state of the tweezer, so the green/dash--dotted line likely overestimates the suppression by combining the density--lowering effects of finite temperature and the strongest possible anti--correlations from a 1D theory. The red/dotted line model in Fig.~\ref{gamma3theory} assumes that the atoms are in the transverse ground state before integrating over transverse dimensions in Eq.~\eqref{eq:corrfunc}. This gives $C_3^3\approx  \frac{9}{4\pi^4 l_\perp^4} \int \mathrm{d}z \,g_3\, n_{1D}(z)^3$, which yields a three-body loss rate that is slightly higher but closer to the observed rate.
While this is predominantly due to the assumed higher atomic density and it does not capture the role played by transversely excited states, 
the red/dotted line makes a prediction for a possible experiment where the atoms are transversely cooled to the ground state.


\begin{figure}[h!]
\centering
\includegraphics[clip, trim=6.6cm 14.6cm 6.3cm 5.6cm,width=0.5\textwidth]{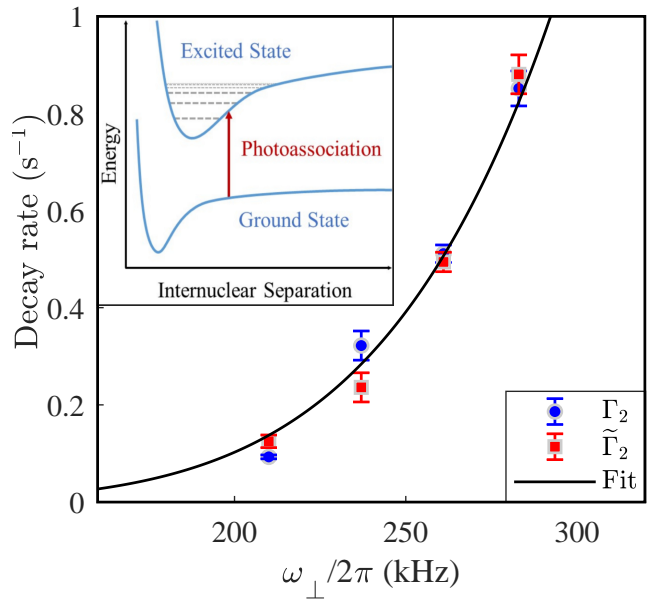}%
\caption{\label{gammaplot}
Measured decay rate using $\widetilde\Gamma_{2}$ for triads (red/square) and $\Gamma_{2}$ for dyads (blue/circle) as a function of trap frequency. \textit{Inset.} illustration of off-resonant photo--association coupling. }
\end{figure}

In addition to the suppressed three--body rate, Fig.\ \ref{decayplot}a) reveals a high pair loss rate (yellow stars). To confirm that this is indeed a two--body loss process, and not a three--body process where only two of the atoms are lost, we utilize our ability to control the initial atomic population. We switch to initial dyad loading [$r_3(t=0)=0$, $r_2(t=0)\approx1$], and obtain population dynamics as in Fig. \ref{decayplot}b).
Fig.\ \ref{gammaplot} shows $\widetilde\Gamma_{2}$ for triad (red squares) and  $\Gamma_{2}$ for dyad (blue circles) loading as a function of transverse trap frequency. Since dyad and triad loading yields $\widetilde\Gamma_2\simeq \Gamma_2$, we conclude that the pair loss observed in Fig.\ \ref{decayplot}a) is the result of a two--body process. 
By preparing single atoms and observing that they remain in the $|F=2\rangle$ ground state for the experiment duration, we can rule out pair loss from the single-frequency tweezer laser causing spontaneous Raman transitions to the $|F=3\rangle$ ground state, followed by a hyperfine changing collision \cite{Xu2015}. Reference \cite{Passagem2017} showed photo--association resonances in the the vicinity of our tweezer wavelength of 1064 nm. To check if our laser frequency coincidentally is at a photo--association resonance we shifted it by 600 MHz, but nevertheless did not observe a significant change in the pair loss rate. 
To investigate whether the observed two--body loss may be 
due to off resonant photo--association \cite{Jones2006} as illustrated in the inset of Fig.\ \ref{gammaplot}, or a multi--photon process to even higher excited two--atom states \cite{Schlagmuller2016}, we fit the data with models that assume that the two-body rate coefficient $K_2=\frac{\Gamma_2}{\int{\mathrm{d^3}r \, n^2\left(\mathbf{r} \right)}}$ is proportional to different integer powers of the tweezer beam intensity. The best fit is obtained assuming a quadratic dependence of $K_2$ on 
the tweezer beam intensity, which gives $\Gamma_2=A \omega_\perp^{11/2}$, with $A$ the fitted parameter \cite{SuppMat}. This model, shown as the solid line, fits the data well, contrary to models assuming that $K_2$ is independent or proportional to the tweezer beam intensity. The observed quadratic dependence of $K_2$ could indicate that the loss involves a two-photon process. For further insight, it would therefore be interesting to change the tweezer beam wavelength in future experiments.

By preparing the atoms with random $m_{F}$-state in the $|F=2\rangle$ manifold we can measure $\Gamma_{3}$ and $\Gamma_{2}$ with effectively distinguishable bosons. The $\Gamma_{3}$ and $\Gamma_{2}$ coefficients reduced by factors of 0.53 and 0.67 respectively. This is consistent with indistinguishable bosons having a statistical tendency to congregate spatially near each other.


Three--body recombination is problematic in many atomic physics experiments as it results in undesired loss events. It is therefore intriguing that we observe the process strongly suppressed.
Our present estimations
indicate that the suppression could be a result of correlations in the multi-particle wave functions due to a combination of  geometric constriction and elastic scattering processes. However, it requires further theoretical developments to verify this, or to clarify whether inelastic three-body processes play a significant role \cite{Daley2009}. Here it is interesting to note that we see evidence of inelastic two-body processes, and these may also lead to anti--correlations that suppress loss \cite{Syassen2008}. An alternative explanation is that the relatively extreme experimental conditions we use alter $\kappa_3$. This could happen due to the strong confinement by the tweezer or the presence of the intense tweezer light. Finally,  strong transverse confinement affects the Efimov physics of three-body bound states \cite{Nishida2018} and resonances, and it is presently unknown where these reside for our trap geometry. While the presence of an Efimov resonance would likely rather enhance the three-body loss rate compared to the background, it is also possible that destructive interference between resonant and non--resonant scattering (Fano effect) could reduce the loss rate, as was observed in Ref.~\cite{Kraemer2006}. 

In conclusion, we present the first study of collisional loss dynamics in individually assembled atomic triads. We confirm that all three atoms are lost in three--body recombination, but observe that the rate of the process is strongly suppressed relative to the rate expected from a thermal sample without interaction--induced correlations. 
Present theory for bulk gases does not fully explain the suppression of three-body loss observed in our experiment but provides a strong indication that interaction-induced anticorrelations cause the effect. Further theoretical developments are needed to understand the dimensional crossover regime that we are probing.
Additionally, the data reveal an unexpected two--body loss process induced by the tweezer laser.
Our approach overcomes the challenge of differentiating between processes faced when trying to infer few--body dynamics from many-body experiments as well as the need for accurate modelling of a time--dependent density profile. It therefore marks a promising direction for future few--body studies, e.g.~the characterisation of Efimov resonances in the dimensional crossover from three to one dimensions by tuning an external magnetic field across a Feshbach resonance. 

We gratefully acknowledge comments on our research from a range of members of the scientific community including J.\ Walraven and D.\ Blume. This work was supported by the Marsden Fund Council from Government funding, administered by the Royal Society of New Zealand (Contracts No.~UOO1835 and MAU1604).  


\bibliography{UOPaperBib}

\begin{widetext}
\section{Supplemental material}
\section{From Master equation to rate equations}
In this work we model atom loss dynamics using the theory of open quantum system in the Born-Markov approximation,
which is adequate if the atoms are completely lost from the trap in processes that happen quickly compared to the time-scale of in-trap dynamics  \cite{Jack2002}. The Born-Markov master equation for the many-body density operator $\hat\rho$ is then given by
\begin{align}
\label{eq:MasterEquation}
    \frac{d\hat\rho}{dt}=&-\frac{i}{\hbar}\left[\hat H_T,\hat\rho\right]+\sum_{j=1}^3\kappa_j\int d^3r\left[2\hat\psi^j(\mathbf r)\hat\rho\hat\psi^{\dagger j}(\mathbf r)-\hat\psi^{\dagger j}(\mathbf r)\hat\psi^j(\mathbf r)\hat\rho-\hat\rho\hat\psi^{\dagger j}(\mathbf r)\hat\psi^j(\mathbf r)\right],
\end{align}
with the non-dissipative part given by the commutator and three Lindblad-type terms for 1,2,3-body losses.\\

To go from here to equations for probabilities, we first separate the density matrix into its $0,1,2,\ldots$-particle components:
\begin{equation}
\label{eq:rhodecomposition}
    \hat\rho=r_0\hat\rho_0+r_1\hat\rho_1+r_2\hat\rho_2+r_3\hat\rho_3+\ldots.
\end{equation}
We define the full density matrix as well as each partial $n$-body component to be normalized, such that
\begin{align}
    \text{Tr}\,\hat\rho &=1, \\
    \text{Tr}\,\hat\rho_i &=1,
\end{align}
and therefore the coefficients must obey
\begin{equation}
    \sum_i r_i=1,
\end{equation}
and can be interpreted as probabilities. We now define the projection operator
\begin{equation}
    \hat P_n=\sum_\nu|\nu,n\rangle\langle \nu,n|, 
\end{equation}
where the sum over $\nu$ denotes a sum over all states in the $n$-body subspace of the Hilbert space. This means, that
\begin{equation}
    \hat P_n\hat\rho=r_n\hat\rho_n
\end{equation}
and $\hat P_n^2=\hat P_n$. The $n$-body subspaces are orthogonal, therefore
\begin{equation}
     \hat P_n\hat\rho_k=\delta_{kn}\hat\rho_k~~ \& ~~\left[\hat P_n,\hat\rho_k\right]=0.
\end{equation}

If we apply this projector to the full Master equation (\ref{eq:MasterEquation}) above, we can split it into equations for each $n$-body component. We obtain
\begin{align}
    \hat P_n \frac{d\hat\rho}{dt}
    &=-\frac{i}{\hbar}r_n\left[H_T,\hat\rho_n\right]+\sum_{j=1}^3\kappa_j\int d^3r\hat P_n\left[2\hat\psi^j\hat\rho\hat\psi^{\dagger j}-\hat\psi^{\dagger j}\hat\psi^j\hat\rho-\hat\rho\hat\psi^{\dagger j}\hat\psi^j\right]\nonumber\\
    &=-\frac{i}{\hbar}r_n\left[H_T,\hat\rho_n\right]+\sum_{j=1}^3\kappa_j\int d^3r\left[2\hat\psi^j\hat\rho_{n+j}r_{n+j}\hat\psi^{\dagger j}-\hat\psi^{\dagger j}\hat\psi^j\hat\rho_n r_n-\hat\rho_n r_n\hat\psi^{\dagger j}\hat\psi^j\right].
\end{align}
Now, we take the trace of this equation. For the left hand side, we obtain
\begin{equation}
    \text{Tr}\,\hat P_n \frac{d\hat\rho}{dt}=\frac{d}{dt}\text{Tr}\,\hat P_n\hat\rho=\frac{d}{dt}\text{Tr}\,r_n\hat\rho_n=\frac{d}{dt}r_n ,
\end{equation}
because of the normalization of $\hat\rho_n$. The non-dissipative term on the right hand side vanishes, since $\text{Tr}[A,B]=0$. What remains is the following expression
\begin{align}
    \frac{d}{dt}r_n=\sum_{j=1}^3 \kappa_j\left\lbrace\text{Tr}\left[2\int d^3r\hat\psi^j\hat\rho_{n+j}\hat\psi^{\dagger j}\right]r_{n+j}-\text{Tr}\left[\int d^3r\hat\psi^{\dagger j}\hat\psi^j\hat\rho_n\right]r_n -r_n\text{Tr}\left[\int d^3r\hat\rho_n\hat\psi^{\dagger j}\hat\psi^j\right]\right\rbrace.
\end{align}
Using the cyclic property of the trace, we can simplify this expression and obtain coupled rate equations for the probabilities $r_n$, 
\begin{equation}
    \frac{d}{dt}r_n=2\sum_{j=1}^3 \kappa_j\left\lbrace r_{n+j}\int d^3x\text{Tr}\left[\hat\psi^{\dagger j}\hat\psi^j\hat\rho_{n+j}\right]-r_n\int d^3x\text{Tr}\left[\hat\psi^{\dagger j}\hat\psi^j\hat\rho_n\right]\right\rbrace.
\end{equation}
The coefficients are given by expectation values of products of field operators in the $n$-body subspace of the Hilbert space, which we denote by
\begin{equation}
    \langle \hat O \rangle_n=\text{Tr}\left(\hat O\hat\rho_n\right).
\end{equation}
With this notation and re-adding the explicit $\mathbf r$-dependence of the fields, the equation above reads as
\begin{equation}
    \frac{d}{dt}r_n=2\sum_{j=1}^3 \kappa_j\left(r_{n+j}\int d^3r\langle\hat\psi^{\dagger j}(\mathbf r)\hat\psi^j(\mathbf r)\rangle_{n+j}-r_n\int d^3r\langle\hat\psi^{\dagger j}(\mathbf r)\hat\psi^j(\mathbf r)\rangle_n\right).
\end{equation}

\subsection{Rate equations for probabilities}\label{sec:rate}
Let us now explicitly write down rate equations for the a 3-body system with just the probabilities $r_n$, $n=0,1,2,3$. First, we have
\begin{align}
\dot r_0&=2\kappa_1\left(r_1\int d^3r\langle\hat\psi^\dagger(\mathbf r)\hat\psi(\mathbf r)\rangle_1-r_0\int d^3r\langle\hat\psi^\dagger(\mathbf r)\hat\psi(\mathbf r)\rangle_0\right)\nonumber\\
&+2\kappa_2\left(r_2\int d^3r\langle\hat\psi^{\dagger 2}(\mathbf r)\hat\psi^2(\mathbf r)\rangle_2-r_0\int d^3r\langle\hat\psi^{\dagger 2}(\mathbf r)\hat\psi^2(\mathbf r)\rangle_0\right)\nonumber\\
&+2\kappa_3\left(r_3\int d^3r\langle\hat\psi^{\dagger 3}(\mathbf r)\hat\psi^3(\mathbf r)\rangle_3-r_0\int d^3r\langle\hat\psi^{\dagger 3}(\mathbf r)\hat\psi^3(\mathbf r)\rangle_0\right),
\end{align}
where we quickly note that the loss terms on the right (with the minus sign) all vanish, because $\langle\hat\psi^{\dagger n}\hat\psi^n\rangle_m=0$ for $m < n$. Also, we use
\begin{equation}
\label{eq:particlenumber}
    \int d^3r \langle\hat\psi^\dagger(\mathbf r)\hat\psi(\mathbf r)\rangle_m=\text{Tr}\int d^3r\hat\psi^\dagger(\mathbf r)\hat\psi(\mathbf r)\hat\rho_m=m\,\text{Tr}\,\rho_m=m.
\end{equation}
to reduce this equation to
\begin{align}
\dot r_0&=2\kappa_1 r_1+2\kappa_2 r_2\int d^3r\langle\hat\psi^{\dagger 2}(\mathbf r)\hat\psi^2(\mathbf r)\rangle_2+2\kappa_3 r_3\int d^3r\langle\hat\psi^{\dagger 3}(\mathbf r)\hat\psi^3(\mathbf r)\rangle_3 .
\end{align}
Analogously, we compute the remaining rate equations for $r_n$, $n=1,2,3$:
\begin{align}
    \dot r_1&=2\kappa_1\left(2r_2-r_1\right)+2\kappa_2 r_3\int d^3r\langle\hat\psi^{\dagger 2}(\mathbf r)\hat\psi^2(\mathbf r)\rangle_3\\
    \dot r_2&=2\kappa_1\left(3r_3-2r_2\right)-r_2\int d^3r\langle\hat\psi^{\dagger 2}(\mathbf r)\hat\psi^2(\mathbf r)\rangle_2\\
    \dot r_3&=2\kappa_1\left(4r_4-3r_3\right)-r_3\int d^3r\langle\hat\psi^{\dagger 2}(\mathbf r)\hat\psi^2(\mathbf r)\rangle_3-r_3\int d^3r\langle\hat\psi^{\dagger 3}(\mathbf r)\hat\psi^3(\mathbf r)\rangle_3.
\end{align}
If we define a shorthand for the integral over the local $n$-body correlation function in the $m$-body subspace
\begin{align} \label{eq:Nmn}
    \mathcal{C}_m^n\equiv \int d^3r\langle\hat\psi^{\dagger n}(\mathbf r)\hat\psi^n(\mathbf r)\rangle_m ,
\end{align}
we obtain the following set of equations:
\begin{align} \label{eq:rateeqs}
\begin{split}
    \dot r_0(t)&=\Gamma_3 r_3+\Gamma_2 r_2+\Gamma_1 r_1 ,\\
    \dot r_1(t)&=-\Gamma_1 r_1+\widetilde\Gamma_2 (3 r_3)+\Gamma_1 (2r_2) ,\\
    \dot r_2(t)&=-\Gamma_2 r_2-\Gamma_1 (2r_2)+\Gamma_1 (3r_3) ,\\
    \dot r_3(t)&=-\Gamma_3 r_3-\widetilde\Gamma_2 (3r_3)-\Gamma_1 (3r_3) ,
\end{split}
\end{align}
where substituting
\begin{align}
      \Gamma_1&=2\kappa_1 ,
\end{align}
yields Eq.\ (2) in the paper and the remaining coefficients are defined as
\begin{align} \label{eq:g1}
\begin{split}
    \Gamma_2&=2\kappa_2 \mathcal{C}_2^2 ,\\
    \widetilde\Gamma_2&=\tfrac23\kappa_2 \mathcal{C}_3^2 ,\\
    \Gamma_3&=2\kappa_3 \mathcal{C}_3^3.
\end{split}
\end{align}

\subsection{Uncorrelated thermal gas}
We can find an expression for the integrated $n$-particle point correlation function $\mathcal{C}_m^n\equiv \int d^3r\langle\hat\psi^{\dagger n}(\mathbf r)\hat\psi^n(\mathbf r)\rangle_m$  of Eq.\ \eqref{eq:Nmn} in terms of the $n$-th power of the single-particle density using the assumptions
\begin{enumerate}
    \item the particles are independently distributed by a given thermal distribution with probability $p_\nu$ to occupy state $|\nu\rangle$,
    \item the probability $p_\nu \ll 1$ for every $\nu$ and thus we may ignore the possibility for more than one particle to occupy the same state.
\end{enumerate}

In this case we may write for the density operator
\begin{align}
    \hat{\rho}_m &\approx \sum_{i_1,\ldots,i_m}p_{i_1}\cdots p_{i_m} a_{i_1}^\dag \cdots a_{i_m}^\dag|\mathrm{vac}\rangle \langle \mathrm{vac}| a_{i_m} \cdots a_{i_1} \\
    &= \left( \hat{\rho}_1 \right)^m .
\end{align}
The correlation function then becomes
\begin{align}
    \langle\hat\psi^{\dagger n}(\mathbf r)\hat\psi^n(\mathbf r)\rangle_m & = 
    \text{Tr} \hat{\rho}_m \hat\psi^{\dagger n}(\mathbf r)\hat\psi^n(\mathbf r) \\
    &=\sum_{i_1,\ldots,i_m}p_{i_1}\cdots p_{i_m} \langle \mathrm{vac}| a_{i_m} \cdots a_{i_1} \hat\psi^{\dagger n}(\mathbf r)\hat\psi^n(\mathbf r) a_{i_1}^\dag \cdots a_{i_m}^\dag|\mathrm{vac}\rangle .
\end{align}
We can resolve the expectation value by commuting the field $\hat\psi(\mathbf r)$ operators past the orbital creation operators towards the right (until they hit the vacuum) using
\begin{align}
    \left[ \hat\psi(\mathbf r), a_{i}^\dag\right] = \phi_i^*(\mathbf r) ,
\end{align}
which give us the value of the orbital function at position $x$.
The $n$ field operators generate $m(m-1)\cdot(m-n+1)$ terms, of which, however, $n!$ terms are identical. The same happens with the field creation operators $\hat\psi^{\dagger}(\mathbf r)$ acting to the left. Only terms with identical content of orbitals will have non-vanishing overlap such that there are $\binom{m}{n}$ terms, but each has a prefactor $(n!)^2$ since it appears on both the left and right hand side. We are thus left with
\begin{align}
    \langle\hat\psi^{\dagger n}(\mathbf r)\hat\psi^n(\mathbf r)\rangle_m & = 
    (n!)^2 \binom{m}{n} \left[\tilde{n}(\mathbf r)\right]^n,
\end{align}
where 
\begin{align}
    \tilde{n}(\mathbf r) = \sum p_i |\phi_i(\mathbf r)|^2 = \frac{n(\mathbf r)}{m} ,
\end{align}
is the single-particle density normalised to 1. 

For the integrated density we thus have 
\begin{align} \label{eq:integrCorr}
   ^\mathrm{th} \mathcal{C}^n_m &= (n!)^2 \binom{m}{n} \int d^3 x \left[\tilde{n}(\mathbf r)\right]^n \\
    & = n! m(m-1) \cdots (m-n+1) \int d^3 x \left[\tilde{n}(\mathbf r)\right]^n \\
    & = \frac{n! m(m-1)\cdots (m-n+1)}{m^n} \int d^3 x \left[n(\mathbf r)\right]^n .
\end{align}
Specifically, for the 3-body term with $n=m=3$ we obtain
\begin{align}
^\mathrm{th}\mathcal{C}^3_3 =\frac43\int d^3r \left[n(\mathbf r)\right]^3 .
\end{align}
The two-particle correlators become
\begin{align}
   ^\mathrm{th} \mathcal{C}^2_3 = 12 \int d^3 x \left[\tilde{n}(\mathbf r)\right]^2 =  3\, ^\mathrm{th}\mathcal{C}^2_2 ,
\end{align}
which implies  $\Gamma_2 =\widetilde\Gamma_2$.

\section{Linear model analytical solution}
The linear loss-rate model [Eq.\ (2) in main paper in the form of Eq.\ \eqref{eq:rateeqs} above] that describes the evolution of the loss events for three body dynamics is:
\begin{equation}
\begin{pmatrix}
    \dot{r_{3}}(t)  \\
    \dot{r_{2}}(t) \\
    \dot{r_{1}}(t) \\
    \dot{r_{0}}(t)
\end{pmatrix} = 
\begin{bmatrix}
    -(\Gamma_{3}+3(\widetilde\Gamma_{2}+\Gamma_{1})) & 0 & 0 & 0 \\
    3\Gamma_{1} & -(\Gamma_{2}+2\Gamma_{1}) & 0 & 0 \\
    3\widetilde\Gamma_{2} & 2\Gamma_{1} & -\Gamma_{1} & 0 \\
    \Gamma_{3} & \Gamma_{2} & \Gamma_{1} &  0
\end{bmatrix}
\begin{pmatrix}
    r_{3}(t)  \\
    r_{2}(t) \\
    r_{1}(t) \\
    r_{0}(t)
\end{pmatrix} .
\end{equation}
The solutions to this system are used to fit to the experimental data in order to extract the rate coefficients $\Gamma_j$.  Their general solutions are:
\begin{equation}
\begin{aligned}
&r_{3}(t)=A\exp[-(\Gamma_{3}+3\widetilde\Gamma_{2}+3\Gamma_{1})t] , \\
&r_{2}(t)=\frac{-3A\Gamma_{1}}{\Gamma_{3}+3\widetilde\Gamma_{2}-\Gamma_{2}+\Gamma_{1}}\exp[-(\Gamma_{3}+3\widetilde\Gamma_{2}+3\Gamma_{1})t]+B\exp[-(\Gamma_{2}+2\Gamma_{1})t] ,\\
&r_{1}(t)=-\alpha\exp[-(\Gamma_{3}+3\widetilde\Gamma_{2}+3\Gamma_{1})t]-\beta\exp[-(\Gamma_{2}+2\Gamma_{1})t]+C\exp(-\Gamma_{1}t) ,\\
&r_{0}(t)=  \left[-\Gamma_{3}A+\left(\frac{3A\Gamma_{1}\Gamma_{2}}{\Gamma_{3}+3\widetilde\Gamma_{2}-\Gamma_{2}+\Gamma_{1}}\right)+\Gamma_{1}\alpha\right]\left(\frac{1}{\Gamma_{3}+3\widetilde\Gamma_{2}+3\Gamma_{1}}\right)\exp[-(\Gamma_{3}+3\widetilde\Gamma_{2}+3\Gamma_{1})t] -C\exp(-\Gamma_{1}t) ,\\
&\qquad+(-\Gamma_{2}B+\Gamma_{1}\beta)\left(\frac{1}{\Gamma_{2}+2\Gamma_{1}}\right)\exp[-(\Gamma_{2}+2\Gamma_{1})t]+D ,
\end{aligned}
\end{equation}
where $~\alpha~\&~\beta$ are:
\begin{equation}
\alpha =  \left(\frac{3A}{\Gamma_{3}+3\widetilde\Gamma_{2}+2\Gamma_{1}}\right)\left[\widetilde\Gamma_{2}-\left(\frac{2\Gamma_{1}^{2}}{\Gamma_{3}+3\widetilde\Gamma_{2}-\Gamma_{2}+\Gamma_{1}}\right)\right]  ;  \qquad \beta =  \left(\frac{2B\Gamma_{1}}{\Gamma_{2}+\Gamma_{1}}\right) ,
\end{equation}
and A, B, C and D are constants of integration evaluated from initial atom populations of zero, one, two, and three atoms in the trap, before collisions began. We estimate these initial populations by analyzing the flourescence collected by the Single Photon Counter Module with $\text{'Wait Time'}=t=0$ as described in the main text. 
The initial populations are:
\begin{equation}
\begin{pmatrix}
    r_{3}(0)  \\
    r_{2}(0) \\
    r_{1}(0) \\
    r_{0}(0)
\end{pmatrix}=
\begin{pmatrix}
    0.836  \\
    0.022 \\
    0.141 \\
    0.001
\end{pmatrix} .
\end{equation}

\section{The two-body loss rate coefficient}
We would like to find $\Gamma_2$'s dependence on $\omega_\perp$ assuming that the two-body loss rate coefficient $K_2$ is proportional to an integer  power ($m$) of the tweezer beam intensity:
\begin{equation} \label{k2}
    \Gamma_2\propto K_2 \int{n^2\left(\mathbf{r} \right) d^3 r} \propto I^m \int{n^2\left(\mathbf{r} \right) d^3 r}.
\end{equation}
Harmonic expansion of the tweezer potential result in $I\propto \omega_\perp^2$, and the density profile of a thermal gas given in the main text gives $\int{n^2\left(\mathbf{r} \right) d^3 r} \propto \omega_\perp^{3/2}$. Insertion into (\ref{k2}) yields:
\begin{equation}
    \Gamma_2\propto \omega_\perp^{2m+3/2}.
\end{equation}

\end{widetext}

\end{document}